\documentclass[conference]{IEEEtran}
\IEEEoverridecommandlockouts
\usepackage{cite}
\usepackage{amsmath,amssymb,amsfonts}
\usepackage{algorithmic}
\usepackage{graphicx}
\usepackage{textcomp}
\usepackage{xcolor}
\usepackage{stfloats}

\def\BibTeX{{\rm B\kern-.05em{\sc i\kern-.025em b}\kern-.08em
    T\kern-.1667em\lower.7ex\hbox{E}\kern-.125emX}}
\begin{document}

\title{Influence of sub-system non-idealities on the performance of Gaussian modulated CV-QKD %system %Sub-systems in Gaussian Modulated CV-QKD: Modelling \& Results
}

\author{\IEEEauthorblockN{R Muralekrishnan\textsuperscript{1}, Lakshmi Narayanan Venkatasubramani\textsuperscript{2}, Sameer Ahmad Mir\textsuperscript{2}, Deepa Venkitesh\textsuperscript{2}}
\IEEEauthorblockA{\textsuperscript{1}\textit{Department of Physics, Indian Institute of Technology Madras,Chennai, India} \\
\textsuperscript{2}\textit{Department of Electrical Engineering, Indian Institute of Technology Madras,Chennai, India} \\
*deepa@ee.iitm.ac.in}
}

\maketitle

\begin{abstract}
We present a detailed analysis of the numerical modelling and evaluation of sub-systems in a Gaussian modulated CV-QKD system, incorporating non-ideal operations, and along with associated results.
\end{abstract}
\vspace{-0.05cm}
\begin{IEEEkeywords}
CV-QKD, Gaussian Modulation, Modelling.
\end{IEEEkeywords}

    \begin{figure*}[b]
        \centering
        \includegraphics[width=01\textwidth]{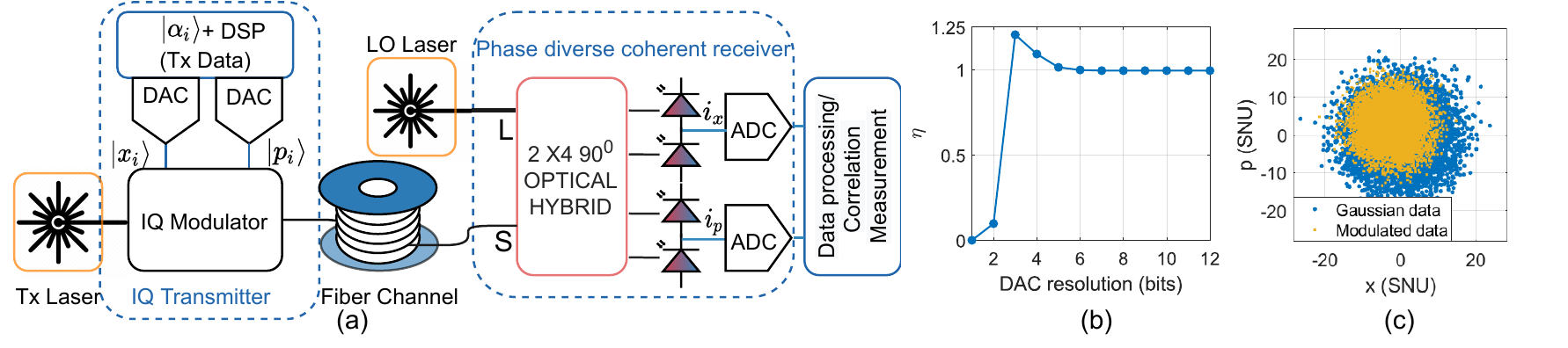}
        \vspace{-0cm}
        \caption{(a) Schematic of the simulation setup for Gaussian modulated CV-QKD (b) Ratio of the measured variance to actual variance ($\eta$) as a function of DAC resolution (c) Constellation after IQ modulator (1 GBd symbol rate) for $\alpha_{IL} = 2$ dB and  $\zeta_{IQmod} = 20$ dB. }
        \label{fig:1}
    \end{figure*}
    \vspace{-0.5cm}
\section{Introduction}
\vspace{-0.2cm}
Quantum Key Distribution (QKD) is a physical layer cryptographic scheme that allows secure information transfer between two authorized users (Alice and Bob), by establishing a secure key, resilient to any eavesdropping (by Eve) [1]. %Depending on the type of information encoding used, the QKD scheme is classified into a discrete variable (DV) or a continuous variable (CV) QKD. 
In continuous variable QKD (CV-QKD), quantum information is encoded in the in-phase and quadrature (IQ) components of the optical field, which are typically modelled as continuous random variables. The CV protocols are convenient to implement due to the ease with which the states can be prepared and measured [1]. Gaussian modulation based CV-QKD protocols such as GG02 [2] are well known and uses components similar to the ones used in classical coherent transmission. However, precise characterization of the subsystems in these protocols are critical for true quantum measurements. In this protocol, Alice prepares coherent states whose quadrature values are drawn from two i.i.d. Gaussian random variables with zero mean and a fixed variance $V_A$ ($|\alpha_i\rangle = |x_i + jp_i\rangle$ where $x_i, p_i \in \mathcal{N}(0, V_A)$). Bob measures the quadrature values using a coherent receiver, which are correlated to the values sent by Alice. Further with reconciliation or reverse reconciliation [3], a secret key is generated that ensures secure communication. Here, we present a numerical analysis of the GG02 protocol, analyzing the effect of non-idealities introduced by each sub-system in the absence of excess noise from Eve.% We observe the effect of different components by comparing the constellations and variance of the data at each step.

\section{Sub-system Models in CV-QKD}

A typical CV-QKD system consists of two highly coherent laser sources (signal and local oscillator (LO)), %In-phase and Quadrature (IQ) 
IQ transmitter for encoding the quantum information in the coherent states of light, fiber channel and phase diverse coherent receiver as shown in Fig. \ref{fig:1}(a). We now discuss in detail the mathematical models used to simulate each component, along with their non-idealities (losses, noise, imperfections). The laser sources in the simulations presented here are assumed to be monochromatic and frequency synchronized.

    \noindent\textbf{(A) Digital to Analog Converter (DAC)}\quad
    \label{DACsection}
        The DAC in the IQ transmitter is used to convert the quadrature values ($x, p$), that are picked from a Gaussian distribution with variance $V_A$, into electrical waveform that drives the IQ modulator. The finite resolution of the DACs limit the precision of the voltage levels, which affects the variance of the data, typically expressed in shot noise units (SNU). If the resolution and peak-to-peak output voltage of the DAC is $n$-bits and $V_{pp}$ respectively, the quantized voltage resolution of the DAC is given by $ V_{step} = \frac{V_{pp}}{2^n-1}$. If a voltage $V$ has to be applied to drive the modulator, the output voltage of the DAC would be approximated to $V_{DAC}= mV_{step} \leq V < (m+1)V_{step}$. The ratio of measured variance to actual variance ($\eta$) is plotted as a function of DAC resolution in Figure \ref{fig:1}(b), which indicates that a minimum of 5-bit resolution is required to truly represent the intended variances. The required output $V_{pp}$ is decided by the IQ modulator and is presented in the next section. Thus it is important to account the quantization noise of the DAC, which depends on $V_{pp}$,  while calculating the $V_A$ at the receiver.

    \noindent\textbf{(B) IQ Modulator}\quad
        The input to the IQ modulator is an electric field of the optical source, represented by the coherent state $|\alpha\rangle$. The DAC output RF voltages, based on the Gaussian data, are applied to the IQ modulator to modulate the optical carrier and there by obtaining the required coherent states $|\frac{\alpha}{2}(\cos\phi_1+j\cos\phi_2)\rangle$. Thus, the two quadrature components in the output field are:
        \begin{equation*}
            x/p = \frac{\alpha}{2}\cos\phi_{x/p}; ~ \phi_{x/p} = \pi\frac{V_{DAC,x/p}}{V_\pi}
            \label{eqn:x-p-quadratures}
        \end{equation*}
        where $\phi_{x/p}$ are related to the voltages ($V_{DAC,x/p}$) that will be applied to the modulator. Therefore, to account for the nonlinear transfer function of the IQ modulator, the voltages required to be applied to the modulator to obtain the required coherent states are
        $V_{DAC,x/p} = \frac{V_\pi}{\pi}\cos^{-1}\left(\frac{x/p}{\alpha/2}\right)$. Insertion loss ($\alpha_{IL}$) of the modulator is modelled as a scaling factor at its output. A finite extinction ratio ($\zeta_{IQmod}$) of the modulator is modelled by introducing additional loss in one of the arms of the Mach Zehnder modulator [4], which causes incomplete destructive interference, leaving a residue. Figure 1(c) shows the shrinking and shift in the constellation due to finite $\alpha_{IL}$ and $\zeta_{IQmod}$ of the modulator respectively. Thus, the input to the DAC needs to be appropriately pre-distorted to compensate for the nonlinearity, finite insertion loss and extinction ratio of the IQ modulator to achieve desired modulation variance.
        \begin{figure*}[t!]
            \centering
            \includegraphics[width=1\textwidth]{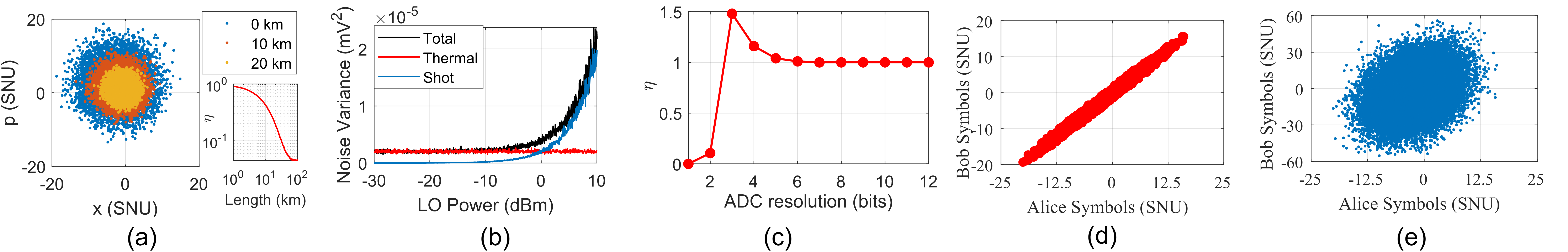}
            \vspace{-0.5cm}
            \caption{(a) Constellation after various lengths of fiber transmission, Variation of $\eta$ with length  (b) Detector noises for various LO power levels (c) Variation of $\eta$ versus ADC resolution (1.2 GHz BW) , and (d) Correlation plot of Tx and Rx coherent states with $P_{LO}=14.1$ dBm and (e) $P_{LO}=-25.9$ dBm.}
            \label{fig:2}
        \end{figure*}
         
    \noindent\textbf{(C) Fiber Channel}\quad
        The fibre channel attenuates ($\alpha$, loss factor) the signal as a function of the transmission length ($l$, in km), leading to a reduction in the variance of the symbols measured by Bob. The loss $L$ in the channel is related to the fibre length as $L=e^{-\alpha l}$. Figure 2(a) shows the constellations plotted at different fiber lengths, and we can observe that the variance of the received signal reduces on increasing the transmission length.

    \noindent\textbf{(D) Phase diverse coherent receiver}\quad
        The receiver contains a $90^o$ hybrid and a balanced receiver to measure the quadrature components of the input electric field. The $90^o$ hybrid combines the signal (S) and the local oscillator (L) fields to give 4 signals that fall on two sets of balanced detectors. After the balanced receiver, we obtain two current components, $i_x= R_DLS\cos\phi$ and $i_p= R_DLS\sin\phi$, that are proportional to the signal quadratures. Here, $R_D$ is responsivity of the detector, and $\phi$ is the phase difference between the signal and the LO fields. Each photodiode adds shot noise and thermal noise to the photocurrent that is generated, modelled as an additive white Gaussian noise with variances $\sigma^2_{shot} = 2eI\Delta f$ and $ \sigma^2_{thermal}= \frac{4KT}{R_L}\Delta f$, respectively. Here, $e$ is the electronic charge, $\Delta f$ is the detector bandwidth, $I$ is the photo-current generated, $K$ is the Boltzmann constant, $R_L$ is the load resistance in the detector and $T$ is the temperature. Figure 2(b) shows noise variance as a function of LO power that allows us to obtain the LO power required to operate the coherent receiver in the shot noise dominated region ($P_{LO} \geq 0~dBm$).

    \noindent\textbf{(E) Analog to Digital Converter (ADC)}\quad 
        The output of the coherent receiver is fed to an ADC to digitize the signal, before performing the correlation measurement. Similar to the discussion in Section \ref{DACsection}(A), the information is quantized to discrete $V_{ADC}$ levels, given as, $ V_{ADC} = (m-\frac{1}{2})V_{step} \leq V < (m+\frac{1}{2})V_{step}.$ The maximum and and minimum values of the ADC are limited by the peak voltage value. Thus, the output voltage of the ADC lies between $-V_{pp}/2$ to $V_{pp}/2$. The ratio of measured variance to actual variance ($\eta$) is plotted as a function of ADC resolution in Figure \ref{fig:2}(c), which indicates that a minimum of 6-bit resolution is required to truly represent the intended variances.%Figure 2(c) shows the output of a finite precision ADC for different input voltages.
         
The correlation plot is a metric to evaluate the robustness of the the coherent state transmission. When our receiver is shot noise dominated (Fig. 2(d)), we are able to observe a better correlation than when the receiver is not shot noise dominated (Fig. 2(e)). The correlation width from the plots reinforce the need to have shot noise-limited operation. For achieving higher secure key rate, the measured correlation needs to be high with a narrow/low spread due to sub-system and excess noises, and it is only possible considering  proper characterization the sub-systems, even in the absence of Eve. 

\section{Result \& Conclusion}
 Using the mathematical models described above, we were able to simulate the Gaussian modulated coherent state CV-QKD protocol and obtain correlations between the transmitted and received data in the presence of noise and non-idealities. 
 
 \small{\textit{ The authors would like to acknowledge funding from Govt. of India agencies: MHRD, SERB and the Office of the PSA and funding from FOCS Project and CQuICC, IIT Madras.}}

\end{document}